\documentclass[longbibliography,aps,prl,reprint,groupedaddress,superscriptaddress]{revtex4-1}

\usepackage[pdftex]{graphicx}
\usepackage{dcolumn}
\usepackage{bm}
\usepackage{amsmath,amssymb}
\usepackage{latexsym}

\usepackage{physics}
\usepackage{verbatim}
\usepackage{lipsum}

\usepackage{xcolor}
\definecolor{mycolor}{rgb}{0.16, 0.17, 0.57}
\usepackage{tikz}

\usepackage[T1]{fontenc}
\usepackage{hyperref}
\hypersetup{colorlinks=true, linkcolor=mycolor, citecolor=mycolor, urlcolor=mycolor,anchorcolor=mycolor}

\usepackage{pdfpages}
\makeatletter
\AtBeginDocument{\let\LS@rot\@undefined}
\makeatother

\renewcommand{\figurename}{FIG.}

\newcommand{\subfiglabel}[1]{\textbf{#1}}

\newcommand{\rfig}[1]{Figure~\textcolor{mycolor}{\ref{#1}}}

\newcommand{\rsubfig}[2]{Fig.~\textcolor{mycolor}{\ref{#1}(#2)}}

\makeindex

\begin{document}
	
	\title{Noise-induced quantum synchronization and maximally entangled  mixed states in superconducting circuits}
    \author{Ziyu Tao}
    \thanks{These authors contributed equally to this work.}
    \affiliation{International Quantum Academy, Futian District, Shenzhen, Guangdong 518048, China}
    \affiliation{Shenzhen Institute for Quantum Science and Engineering and Department of Physics, Southern University of Science and Technology, Shenzhen 518055, China}
    \affiliation{Guangdong Provincial Key Laboratory of Quantum Science and Engineering,
    Southern University of Science and Technology, Shenzhen 518055, China}
    \affiliation{Shenzhen Key Laboratory of Quantum Science and Engineering, Southern University of Science and Technology, Shenzhen,518055, China}

    \author{Finn Schmolke}
    \thanks{These authors contributed equally to this work.}
    \affiliation{Institute for Theoretical Physics I, University of Stuttgart, D-70550 Stuttgart, Germany}

    \author{Chang-Kang Hu}
    \thanks{These authors contributed equally to this work.}
    \affiliation{International Quantum Academy, Futian District, Shenzhen, Guangdong 518048, China}
    \affiliation{Shenzhen Institute for Quantum Science and Engineering and Department of Physics, Southern University of Science and Technology, Shenzhen 518055, China}
    \affiliation{Guangdong Provincial Key Laboratory of Quantum Science and Engineering,
    Southern University of Science and Technology, Shenzhen 518055, China}
    \affiliation{Shenzhen Key Laboratory of Quantum Science and Engineering, Southern University of Science and Technology, Shenzhen,518055, China}

    \author{Wenhui Huang}
    \affiliation{Shenzhen Institute for Quantum Science and Engineering and Department of Physics, Southern University of Science and Technology, Shenzhen 518055, China}
    \affiliation{Guangdong Provincial Key Laboratory of Quantum Science and Engineering,
    Southern University of Science and Technology, Shenzhen 518055, China}
    \affiliation{Shenzhen Key Laboratory of Quantum Science and Engineering, Southern University of Science and Technology, Shenzhen,518055, China}

    \author{Yuxuan Zhou}
    \affiliation{International Quantum Academy, Futian District, Shenzhen, Guangdong 518048, China}
    \affiliation{Shenzhen Institute for Quantum Science and Engineering and Department of Physics, Southern University of Science and Technology, Shenzhen 518055, China}
    \affiliation{Guangdong Provincial Key Laboratory of Quantum Science and Engineering,
    Southern University of Science and Technology, Shenzhen 518055, China}
    \affiliation{Shenzhen Key Laboratory of Quantum Science and Engineering, Southern University of Science and Technology, Shenzhen,518055, China}

    \author{Jiawei Zhang}
    \affiliation{Shenzhen Institute for Quantum Science and Engineering and Department of Physics, Southern University of Science and Technology, Shenzhen 518055, China}
    \affiliation{Guangdong Provincial Key Laboratory of Quantum Science and Engineering,
    Southern University of Science and Technology, Shenzhen 518055, China}
    \affiliation{Shenzhen Key Laboratory of Quantum Science and Engineering, Southern University of Science and Technology, Shenzhen,518055, China}

    \author{Ji Chu}
    \affiliation{International Quantum Academy, Futian District, Shenzhen, Guangdong 518048, China}
    \affiliation{Shenzhen Institute for Quantum Science and Engineering and Department of Physics, Southern University of Science and Technology, Shenzhen 518055, China}
    \affiliation{Guangdong Provincial Key Laboratory of Quantum Science and Engineering,
    Southern University of Science and Technology, Shenzhen 518055, China}
    \affiliation{Shenzhen Key Laboratory of Quantum Science and Engineering, Southern University of Science and Technology, Shenzhen,518055, China}

    \author{Libo Zhang}
    \affiliation{Shenzhen Institute for Quantum Science and Engineering and Department of Physics, Southern University of Science and Technology, Shenzhen 518055, China}
    \affiliation{Guangdong Provincial Key Laboratory of Quantum Science and Engineering,
    Southern University of Science and Technology, Shenzhen 518055, China}
    \affiliation{Shenzhen Key Laboratory of Quantum Science and Engineering, Southern University of Science and Technology, Shenzhen,518055, China}

    \author{Xuandong Sun}
    \affiliation{Shenzhen Institute for Quantum Science and Engineering and Department of Physics, Southern University of Science and Technology, Shenzhen 518055, China}
    \affiliation{Guangdong Provincial Key Laboratory of Quantum Science and Engineering,
    Southern University of Science and Technology, Shenzhen 518055, China}
    \affiliation{Shenzhen Key Laboratory of Quantum Science and Engineering, Southern University of Science and Technology, Shenzhen,518055, China}

    \author{Zecheng Guo}
    \affiliation{Shenzhen Institute for Quantum Science and Engineering and Department of Physics, Southern University of Science and Technology, Shenzhen 518055, China}
    \affiliation{Guangdong Provincial Key Laboratory of Quantum Science and Engineering,
    Southern University of Science and Technology, Shenzhen 518055, China}
    \affiliation{Shenzhen Key Laboratory of Quantum Science and Engineering, Southern University of Science and Technology, Shenzhen,518055, China}

    \author{Jingjing Niu}
    \affiliation{International Quantum Academy, Futian District, Shenzhen, Guangdong 518048, China}

    \author{Wenle Weng}
    \affiliation{Institute for Photonics and Advanced Sensing (IPAS) and School of Physics, Chemistry and Earth Sciences, The University of Adelaide, Adelaide, South Australia 5005, Australia}

    \author{Song Liu}
    \affiliation{International Quantum Academy, Futian District, Shenzhen, Guangdong 518048, China}
    \affiliation{Shenzhen Institute for Quantum Science and Engineering and Department of Physics, Southern University of Science and Technology, Shenzhen 518055, China}
    \affiliation{Guangdong Provincial Key Laboratory of Quantum Science and Engineering,
    Southern University of Science and Technology, Shenzhen 518055, China}
    \affiliation{Shenzhen Key Laboratory of Quantum Science and Engineering, Southern University of Science and Technology, Shenzhen,518055, China}

    \author{Youpeng Zhong}
    \email{zhongyp@sustecch.edu.cn}
    \affiliation{International Quantum Academy, Futian District, Shenzhen, Guangdong 518048, China}
    \affiliation{Shenzhen Institute for Quantum Science and Engineering and Department of Physics, Southern University of Science and Technology, Shenzhen 518055, China}
    \affiliation{Guangdong Provincial Key Laboratory of Quantum Science and Engineering,
    Southern University of Science and Technology, Shenzhen 518055, China}
    \affiliation{Shenzhen Key Laboratory of Quantum Science and Engineering, Southern University of Science and Technology, Shenzhen,518055, China}

    \author{Dian Tan}
    \email{tand@sustech.edu.cn}
    \affiliation{International Quantum Academy, Futian District, Shenzhen, Guangdong 518048, China}
    \affiliation{Shenzhen Institute for Quantum Science and Engineering and Department of Physics, Southern University of Science and Technology, Shenzhen 518055, China}
    \affiliation{Guangdong Provincial Key Laboratory of Quantum Science and Engineering,
    Southern University of Science and Technology, Shenzhen 518055, China}
    \affiliation{Shenzhen Key Laboratory of Quantum Science and Engineering, Southern University of Science and Technology, Shenzhen,518055, China}

    \author{Dapeng Yu}
    \email{yudp@sustech.edu.cn}
    \affiliation{International Quantum Academy, Futian District, Shenzhen, Guangdong 518048, China}
    \affiliation{Shenzhen Institute for Quantum Science and Engineering and Department of Physics, Southern University of Science and Technology, Shenzhen 518055, China}
    \affiliation{Guangdong Provincial Key Laboratory of Quantum Science and Engineering,
    Southern University of Science and Technology, Shenzhen 518055, China}
    \affiliation{Shenzhen Key Laboratory of Quantum Science and Engineering, Southern University of Science and Technology, Shenzhen,518055, China}
    
    \author{Eric Lutz}
    \email{eric.lutz@itp1.uni-stuttgart.de}
    \affiliation{Institute for Theoretical Physics I, University of Stuttgart, D-70550 Stuttgart, Germany}

\begin{abstract}
    Random fluctuations can lead to cooperative effects in complex  systems. We here report the experimental observation of noise-induced quantum synchronization in  a chain of superconducting transmon qubits with nearest-neighbor interactions. The application of Gaussian white noise to a single site leads to synchronous oscillations in the entire chain. We show that the two synchronized end qubits are entangled, with nonzero concurrence, and that they belong to a class of generalized  Bell states known as maximally entangled mixed states, whose entanglement cannot be
increased by any global unitary. We further demonstrate the stability against frequency detuning of both synchronization and entanglement by determining the corresponding generalized Arnold tongue diagrams. Our results highlight the constructive influence of noise in a quantum many-body system and uncover the potential role of synchronization for mixed-state quantum information science.
\end{abstract}

\maketitle
Noise is widely regarded as a nuisance that limits the transmission and processing of information \cite{don62}. The adverse effect of random fluctuations is even more dramatic in quantum physics, since quantum coherence and quantum correlations, two essential resources of quantum technologies \cite{nie00}, are highly susceptible to external disturbances \cite{zur03}. Surprisingly, the nontrivial interplay between noise and nonlinear dynamics may induce order and organization \cite{dix02,ber05,don08,rid11}.  Classical noise-induced phenomena, such as pattern formation, noise-induced transport and stochastic resonance,  occur in  a great variety of contexts, from physics and chemistry to biology and engineering \cite{dix02,ber05,don08,rid11}. However, observing the constructive role of noise   in quantum systems is far \mbox{more challenging \cite{kra11,Wagner2019,Haenze2021,vic15,pot18,Maier2019}.}

Noise-induced synchronization is another counterintuitive consequence of random fluctuations \cite{Zhou2002,ter04,zho05,Nakao2007,lai11,uch04,sun14,Neimann2002,ter08,tou20}; it is, for instance, thought to be relevant for collective neuronal synchronization in the brain \cite{Neimann2002,ter08,tou20}. Synchronization is a general concept in classical \cite{ble88,Boccaletti2002,mos02,ace05,pik03,ani07,Balanov2009} and quantum \cite{Lee2013,Mari2013,Walter2014,Loerch2017,Sonar2018,Roulet2018,Cabot2019,Laskar2020,Buca2022,Schmolke2022_main} physics: synchronous motion usually arises when coupled nonlinear oscillators adjust their internal rhythms, and oscillate in unison \cite{ble88,Boccaletti2002,mos02,ace05,pik03,ani07,Balanov2009,Lee2013,Mari2013,Walter2014,Loerch2017,Sonar2018,Roulet2018,Cabot2019,Laskar2020,Buca2022,Schmolke2022_main}. Synchronization phenomena have lately found interesting applications in communication systems \cite{arg05,choi17}. In noise-induced synchronization, collective oscillations arise through the constructive influence of a noise source. This classical effect has  been observed in lasers \cite{uch04,sun14} and in sensory neurons \cite{Neimann2002}. In the quantum regime, noise-induced synchronization has been predicted to occur in many-body systems \cite{Schmolke2022_main}, but  has not been experimentally observed so far.

We here report the experimental realization of noise-induced quantum synchronization in a linear chain of superconducting transmon qubits with nearest-neighbor interactions \cite{kja20}. We observe the occurrence of stable, synchronized oscillations  of the magnetizations of the edge qubits when Gaussian noise is applied to a single qubit. We further show that the corresponding synchronized state is not only entangled, with nonzero concurrence \cite{woo98}, but that it is given by a maximally entangled  mixed state which exhibits the maximum obtainable amount of entanglement for a given degree of mixedness \cite{ish00,mun01,wei03,pet04,bar04,chi11}. Such states are regarded as direct generalizations of maximally entangled pure Bell states \cite{ish00,mun01,wei03,pet04,bar04,chi11}, and play an important  role in mixed-state quantum information processing \cite{ben96,hor98,aha98,amb06}. We additionally confirm the  robustness of the observed quantum synchronization phenomenon to detuning of the natural frequencies of the qubits.
We concretely  obtain Arnold-tongue-like patterns \cite{ble88,Boccaletti2002,mos02,ace05,pik03,ani07,Balanov2009}, for both synchronization and entanglement, as a function of detuning and  noise strength.

\begin{figure}[t]
	\centering
	\includegraphics[width=.48\textwidth]{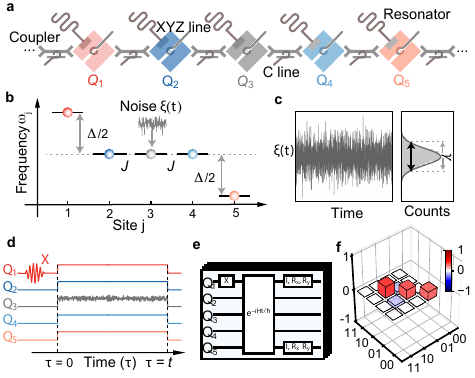}
	\caption{\label{fig_pulse}
	 Experimental system. \subfiglabel{a} The chip of a transmon qubit chain  is designed with a flip-chip technique  with the top layer consisting of tunable-frequency qubits and tunable-frequency couplers, where the couplers facilitate interactions between qubits. Other elements of the chip sit in the bottom layer including the resonators, the qubit control and flux lines ($XYZ$ lines) and neighboring coupler flux lines ($C$ lines). \subfiglabel{b} A one-dimensional $XY$ chain of  five transmon qubits, with coupling constant $J$ and edge-spin frequencies detuned by $\Delta$,  is subjected to \subfiglabel{c} Gaussian white noise $\xi(t)$ with effective amplitude $\gamma$. \subfiglabel{d} The system is initially prepared in a separable state where qubit $Q_1$ is prepared in the excited state while all the other qubits $Q_j$ of the quantum chain are in the ground state; external noise $\xi(t)$ is applied to the central spin $Q_3$. \subfiglabel{e} Quantum circuits used in the tomographic measurement of the two-qubit density matrix $\rho_{15}$ of the two end spins. \subfiglabel{f} The measured real part of the density matrix $\rho_{15}$ of the synchronized edge qubits at time $Jt = 3\pi$ corresponds to a maximally entangled mixed state.
	}
\end{figure}

\textit{Experimental system.} Our experiments are implemented on a superconducting quantum processor, comprising a one-dimensional  array of tunably coupled transmon qubits \cite{kja20} (Figs.~1a,b).
The qubits act as artificial spins, where the $j$th qubit frequency $\omega_j/(2\pi)$ can be controlled, in the range from  $\sim$3.2 GHz to $\sim$4.6 GHz, by applying an external flux through the dedicated $Z$ line  of the corresponding qubit and coupler~\cite{Xu2020a}. Likewise, the coupling constants $(J_j=J)/(2\pi)$ between the qubits is set to  $\sim$10 MHz   by applying an external flux on the associated $C$ line~\cite{Xu2020a} (Supplemental Information).
Each qubit can be individually addressed and driven into the excited state by applying
a microwave pulse through its $XY$ control line. 
The lattice model of the experiments can be described by the Hamiltonian of a one-dimensional quantum $XY$ chain of $N$ spins in a transverse field  \cite{tak99}
\begin{equation}
\label{1}
	H_0 = \frac{\hbar J}{2} \sum_{j=1}^{N-1} 
	\left(\sigma_j^x \sigma_{j+1}^x + \sigma_j^y \sigma_{j+1}^y\right)
	+  \sum_{j=1}^N \hbar \omega_j\sigma_j^z,
\end{equation}
where $\sigma_j^{x,y,z}$ are the local Pauli operators acting on site $j$. We apply Gaussian white noise $\xi(t)$ with zero mean  and autocorrelation $\langle \xi(t) \xi(t^\prime) \rangle = \Gamma \delta(t-t^\prime)$, with noise strength $\Gamma$,  by locally modulating the  natural frequencies of the individual qubits on the desired sites~\cite{Averin2016} (Supplemental Information) (Fig.~1c). This stochastic contribution corresponds to the addition of the Hermitian operator $ \xi(t) \sigma^z_u$ (acting locally on site $u$) to the Hamiltonian \eqref{1}.
Stable synchronization of the entire quantum chain is predicted to occur, for a homogeneous chain with constant frequencies, $\omega_j = \omega$, and noise applied to a single site $u$, when the two conditions,   $N = 5+3m$ and $ u = 3n$ ($m,n \in \mathbb{N}$), are satisfied \cite{Schmolke2022_main}. The latter condition ensures that all the eigenmodes of the many-body system exponentially decay to zero, except one. This leads to a decoherence-free subspace, that is decoupled from the surroundings \cite{lid98}, with only a single eigenmode, whose frequency determines the synchronization frequency. In the following, we analyze the occurrence of stable noise-induced quantum synchronization in a chain of $N=5$ qubits (corresponding to $m=0$) when the noise acts on the third qubit (corresponding to $n=1$) in the middle of the chain (Fig.~1d). This is the smallest spin chain in which noise-induced quantum synchronization is expected to appear. Additional cases with larger number of spins ($N= 8$ and 11), as well as with violations of the synchronization conditions are presented in the \mbox{Supplemental Information}.

\begin{figure}[t!]
	\centering
	\includegraphics[width=0.48\textwidth]{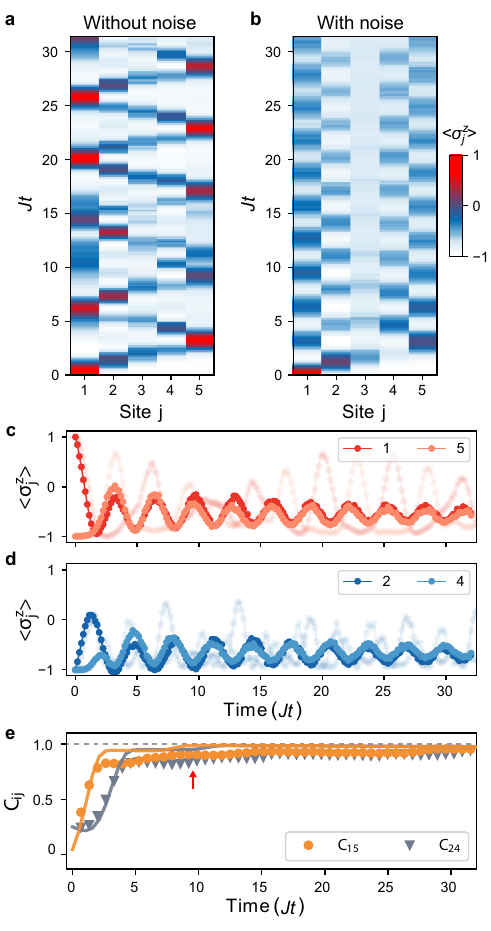}
	\caption{\label{fig_5qEvo} Noise-induced quantum synchronization. \subfiglabel{a} The local $z$-polarizations $\langle \sigma^z_j\rangle$ of a chain of five qubits do not synchronize during noise-free unitary dynamics. \subfiglabel{b} Synchronized oscillations, with frequency $\sim$$ 2J = 20$ MHz, between \subfiglabel{c} qubits 1-5 and \subfiglabel{d} qubits 2-4 occur, when Gaussian white noise with strength $\gamma = 1.3$ is applied to the third spin. The associated unsynchronized unitary evolution is shown in the background. 
	\subfiglabel{e} Pearson correlation coefficients $C_{15}$ and $C_{24}$ of qubits 1-5 and qubits 2-4 (symbols)  converge to  one, indicating perfect correlation between the corresponding qubits. The synchronized regime, $C_{ij}\geq 0.9$, is attained for $Jt\geq 3\pi$ (red arrow). Good agreement with theory (solid lines) is obtained.
	}
\end{figure}

\textit{Noise-induced quantum synchronization.} We initially prepare the quantum chain in the separable state $\ket{\Psi(0)} = \ket{1} \otimes \ket{0}^{\otimes 4}$, where $\ket{1}$ and $\ket{0}$ denote the respective excited and  ground states of the qubits (Fig.~1d).
The state of each qubit is determined by measuring the state-dependent transmission of a dedicated readout resonator, with frequency around 6.1 GHz,  coupled to each qubit using a dispersive readout scheme (Supplemental Information). \rfig{fig_5qEvo} displays  the temporal evolution of the measured local $z$-polarization $\langle\sigma^z_j\rangle$ of the individual qubits, without (Fig.~\ref{fig_5qEvo}a) and with (Fig.~\ref{fig_5qEvo}b) Gaussian noise applied to the middle of the chain (the reduced noise strength is $\gamma=\Gamma/J \approx 1.3$).
In the absence of noise, the initial excitation travels through the chain in a wave-like manner, is reflected at the open boundaries, and bounces back and forth between the edges of the chain, without collective coordination between the individual qubits (Fig.~\ref{fig_5qEvo}a).
By contrast, in the presence of Gaussian noise, qubits 1 and 5 (as well as qubits 2 and 4) oscillate in phase at a single  frequency of $\sim$20 MHz, in agreement with the predicted value $2J$, after some transient  (Figs.~\ref{fig_5qEvo}b,c). The synchronization frequency is set by the coupling constant of the quantum  chain and not by the eigenfrequencies of the qubits. The faint lines in the background of Figs.~\ref{fig_5qEvo}c,d represent the measured unsynchronized, quasi-unitary dynamics of the magnetizations, when the  external Gaussian noise is switched off.

In order to quantitatively characterize the  synchronized oscillations  of the respective polarizations,  we use the Pearson correlation coefficient,  defined as the ratio of
the covariance and the respective standard deviations, $C_{ij}=\mathrm{cov}\left(\langle \sigma^z_i\rangle,\langle \sigma^z_j\rangle\right)/\sqrt{\mathrm{var}(\langle \sigma^z_i\rangle) \mathrm{var}(\langle \sigma^z_j\rangle)}$ \cite{bar89}. The latter quantity provides a measure of the degree of linear correlation between observables; it ranges from $-1$ (corresponding to anticorrelated oscillations) to $+1$ (indicating correlated evolution).
Figure 2e shows the Pearson correlation coefficients $C_{15}$ and $C_{24}$ of the measured $z$-polarizations of qubits 1-5 and qubits 2-4, as a function of time (symbols).
Both  quickly converge to  one, demonstrating almost perfectly correlated oscillations over the entire duration of the experiment, in good agreement with the theoretical simulations (solid lines) (Supplemental Information). We specifically consider two qubits to be synchronized for Pearson coefficients $C_{ij}\geq 0.9$, which happens  for times $Jt\geq 3\pi$ (red arrow).

\begin{figure*}[t!]
	\centering
	\includegraphics[width=0.9\textwidth]{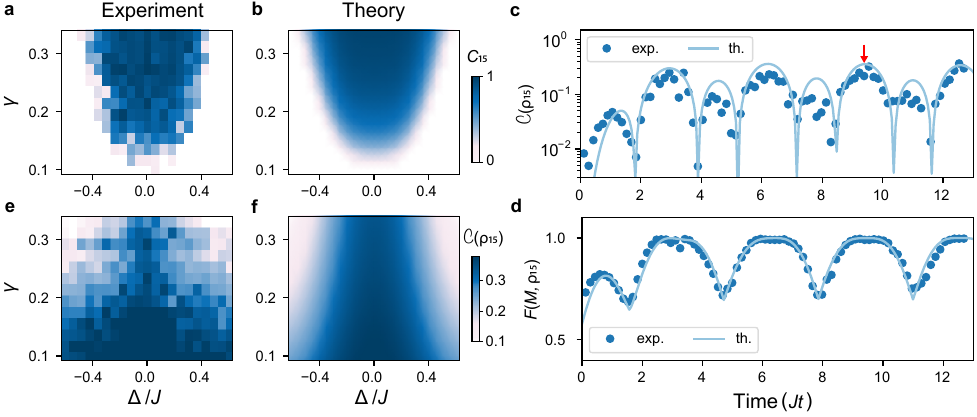}
	\caption{\label{fig_tongue}
		Stability regions and entanglement.
            \subfiglabel{a}-\subfiglabel{b}
        Arnold tongue of synchronization.
        Experimental data (\subfiglabel{a}) and numerical simulation (\subfiglabel{b}) for the Pearson correlation coefficients $C_{15}$, 
        extracted at a given time $Jt = 3\pi$ (red arrow), as a function of noise amplitude $\gamma$ and detuning $\Delta$ between the two end spins.
        Larger values of detuning are detrimental, but the system can still be synchronized by increasing the noise strength.
        \subfiglabel{c}-\subfiglabel{d},
        Concurrence $\mathcal{C}(\rho_{15})$ (\subfiglabel{c}) and fidelity $F(M,\rho_{15})$ (\subfiglabel{d})
        of the predicted maximally entangled mixed state ($M$) and the measured  two-qubit density operator  ($\rho_{15}$) as a function of time. Both quantities display non-vanishing steady oscillations for $Jt\geq 2\pi$, showing that the synchronized two-qubit state is entangled
        Good agreement between data (dots) and theory (lines) is observed.        \subfiglabel{e}-\subfiglabel{f}
        Entanglement tongue.
        Experimental data (\subfiglabel{e}) and numerical simulation (\subfiglabel{f}) for the concurrence $\mathcal{C}(\rho_{15})$ at a given time $Jt = 3\pi$ as a function of noise amplitude $\gamma$ and the detuning $\Delta$.
        Increasing both the detuning and the noise strength diminishes the amount of entanglement in the system.}
        \end{figure*}

We next analyze  the robustness of  the synchronous oscillations. To that end,
we introduce  a variable detuning $\Delta$ between the natural frequencies of the synchronized end qubits via  a term $H_1 = (\hbar\Delta/2)(\sigma_1^z - \sigma^z_5)$ added to the system Hamiltonian \eqref{1} (Fig.~1b). Figure \ref{fig_tongue}a displays the measured Pearson correlation coefficient  $C_{15}$ at the onset of the synchronization regime
$Jt=3\pi$, when  both the reduced noise amplitude $\gamma$ and the detuning $\Delta$ are varied. We recognize a structure which is reminiscent of an Arnold tongue which defines the synchronized domain of classical synchronization  phenomena \cite{ble88,Boccaletti2002,mos02,ace05,pik03,ani07,Balanov2009}. Noise-induced quantum synchronization appears to be a robust effect that occurs in a wide range of parameters of the system. We note that  the synchronization region is  enlarged when the detuning is reduced and when the noise strength is increased; it is thus easier to synchronize identical spins with strong noise (provided the noise amplitude remains below the noise-induced quantum Zeno  regime \cite{whi09,bur20}). We again obtain good agreement with theoretical simulations (Fig.~\ref{fig_tongue}b).

\textit{Maximally entangled mixed states.}
Entanglement is a fundamental resource in quantum information science; mechanisms creating entangled states are hence of great importance \cite{nie00}.
There seems, however,  to be no direct relationship between quantum synchronization and  quantum correlations in general \cite{kar19,lia19,ste23}. In view of the detrimental influence of noise on quantum properties \cite{zur03}, it is therefore all the more remarkable that noise-induced synchronization is expected to give rise to entangled synchronized edge qubits \cite{Schmolke2022_main}. In order to test this feature, we tomographically reconstruct the two-qubit state of the two end spins \cite{par04} as illustrated in Fig.~1e and evaluate the concurrence, $\mathcal{C}(\rho_{15}) = \text{max}\left(0,\sqrt{\kappa_1}-\sqrt{\kappa_2}-\sqrt{\kappa_3}-\sqrt{\kappa_3}\right)$; the operator $\rho_{15} $ is here  the reduced density matrix of the two edge qubits and {$\kappa_n$} are the ordered eigenvalues of the product $\rho_{15} \widetilde \rho_{15}$, with  $\widetilde \rho_{15}$   the spin flipped state \cite{woo98}. Figure~\ref{fig_tongue}c shows that $\mathcal{C}(\rho_{15})$ exhibits nonzero steady oscillations, clearly indicating the presence of  synchronized  entangled edge qubits. This observation reveals an intriguing connection between collective quantum behavior and nonclassical correlations. The concurrence reaches a steady  state for $Jt\geq 2\pi$, thus slightly before the two edge qubits are fully synchronized ($Jt\geq 3\pi$).

The  reconstructed two-qubit state $\rho_{15}$ at time $Jt = 3\pi$ is explicitly given in Fig.~1f: it has the form of a maximally entangled mixed state which can be parametrized as $\rho = p_1 \dyad{\Psi^-} + p_2 \dyad{00} + p_3 \dyad{\Psi^+} + p_4 \dyad{11}$ \cite{ish00},
 where $\ket{\Psi^-} = (\ket{01} - \ket{10})/\sqrt{2}$ and $\ket{\Psi^+} = (\ket{01} + \ket{10})/\sqrt{2}$ are the usual Bell states \cite{nie00} and  $\sum_k p_k = 1$. The concurrence of the maximally entangled mixed state can be analytically determined as $\mathcal{C} = p_1-p_3-2\sqrt{p_2p_4}$ \cite{ish00}. The fidelity of the measured state $\rho_{15}$ and the theoretical maximally entangled mixed state $M$ given, for the parameters of the experiment, by 
 \begin{align}
    M =     
    \begin{pmatrix}
    0 & 0 & 0 & 0 \\
    0 & 1/3 & -1/6 & 0 \\
    0 & -1/6 & 1/3 & 0 \\
    0 & 0 & 0 & 1/3
    \end{pmatrix},
\end{align}
is $F(M,\rho_{15})= \operatorname{Tr} \big[{ \scriptstyle \sqrt{\sqrt{M} \rho_{15}\sqrt{M}} }\big] = 99.3\%$ (Supplemental Information). The maximally entangled mixed state is hence already created at the beginning of the synchronized regime (the fidelity between measured and theoretical states at $Jt= 4\pi$ is $F=99.6\%$).  The time evolution of the fidelity of the measured two-qubit state and the theoretical maximally entangled mixed state is shown in Fig.~\ref{fig_tongue}d. It exhibits a behavior  similar to that of the concurrence (Fig.~\ref{fig_tongue}c). In particular, it displays steady oscillations for $Jt\geq 2\pi$. Maximally entangled mixed states define a class of quantum states for which no more entanglement can be created by any global unitary operations  \cite{ish00,mun01,wei03,pet04,bar04,chi11}. They have the interesting property that they are more entangled than Werner states with the same  purity \cite{ish00,mun01,wei03,pet04,bar04,chi11}. So far, maximally entangled mixed states have only been generated in optical systems \cite{pet04,bar04,chi11}.

Like the Pearson correlation coefficient $C_{15}$, the concurrence $\mathcal{C}(\rho_{15})$ is robust to detuning of the edge qubit frequencies. It appears in a large parameter domain (Fig.~\ref{fig_tongue}e), and  exhibits an (inverted) Arnold-tongue-like structure  that  results from the competition of two different mechanisms: on the one hand, entanglement between the initially separable end spins is created through the unitary evolution of  the system  \cite{Zanardi2000,Zanardi2001,Konrad2008,Linden2009}; on the other hand, noise, which drives the quantum synchronization process, destroys quantum correlations. The value of the effective noise strength $\gamma$, which controls the synchronization time \cite{Schmolke2022_main}, sets the maximal amount of entanglement that the synchronized state can have once it has reached the (quasi)-stationary state in the decoherence-free subspace. The concurrence of the maximally entangled mixed state thus decreases when the noise amplitude or the detuning are increased. The entanglement tongue seen in Fig.~\ref{fig_tongue}e is often regarded as a quantum generalization of the classical Arnold tongue \cite{Lee2014}. As before, good agreement with theoretical simulations is found (Fig.~\ref{fig_tongue}f).

\textit{Conclusions.} We have experimentally demonstrated the occurrence of  noise-induced quantum synchronization in a chain of transmon qubits by applying Gaussian white noise to {one site of the chain}. Perfectly correlated in-phase oscillations at a frequency set by the coupling constant of the chain, with a Pearson   coefficient close to one,  have been observed. These findings provide a unique illustration of the nontrivial interplay between noise and unitary dynamics in a quantum many-body system, leading to collective behavior, and at the same time, to the creation of distant quantum correlations with nonzero concurrence. In view of the importance of spin chains for solid-state quantum information processing \cite{bos03,chr04,fit06,kay07,fra08,nor12,sah15}, we expect the generation of maximally entangled mixed states in these systems to be relevant  for synchronization-based quantum communications \cite{arg05,choi17}.

\textit{Acknowledgments.—}
This work is supported by the Key-Area Research and Development Program of Guangdong Province (2018B030326001), the National Natural Science Foundation of China (12004167, 11934010, 12174178, 12075110, 12374474, 12205137), the National Key Research and Development Program of China (2019YFA0308100), the Guangdong Innovative and Entrepreneurial Research Team Program (2016ZT06D348, 2019ZT08C044), the Guangdong Provincial Key Laboratory (2019B121203002), the Science, Technology and Innovation Commission of Shenzhen Municipality (KYTD PT20181011104202253, KQTD20210811090049034, KQTD20190929173815000, JCYJ20200109140803865, KQTD20210811090049034), the Shenzhen-Hong Kong Cooperation Zone for Technology and Innovation (HZQB- KCZYB-2020050), the NSF of Beijing (Z190012) and the Guangdong Basic and Applied Basic Research Foundation (2022A1515110615), and the Innovation Program for Quantum Science and Technology (2021ZD0301703). F.S and E. L. acknowledge financial support from the Vector Foundation and the DFG (Grant FOR 2724).


%

\clearpage

\widetext

\renewcommand{\bibnumfmt}[1]{[S#1]}
\renewcommand{\citenumfont}[1]{S#1}
\setcounter{equation}{0}
\setcounter{figure}{0}
\setcounter{table}{0} 
\setcounter{page}{1}

\renewcommand{\theequation}{S\arabic{equation}}
\renewcommand{\thefigure}{S\arabic{figure}}
\renewcommand{\bibnumfmt}[1]{[S#1]}
\renewcommand{\citenumfont}[1]{S#1}

\renewcommand{\figurename}{Supplementary Fig.}
\renewcommand{\tablename}{Supplementary Table}

\newcommand{\figtitle}[1]{{#1}}
\newcommand{\subcaptionfont}{\fontfamily{phv}\selectfont}
\newcommand{\subcaptionlabel}[1]{\subcaptionfont({#1})}

\newpage



\includepdf[pages={1}]{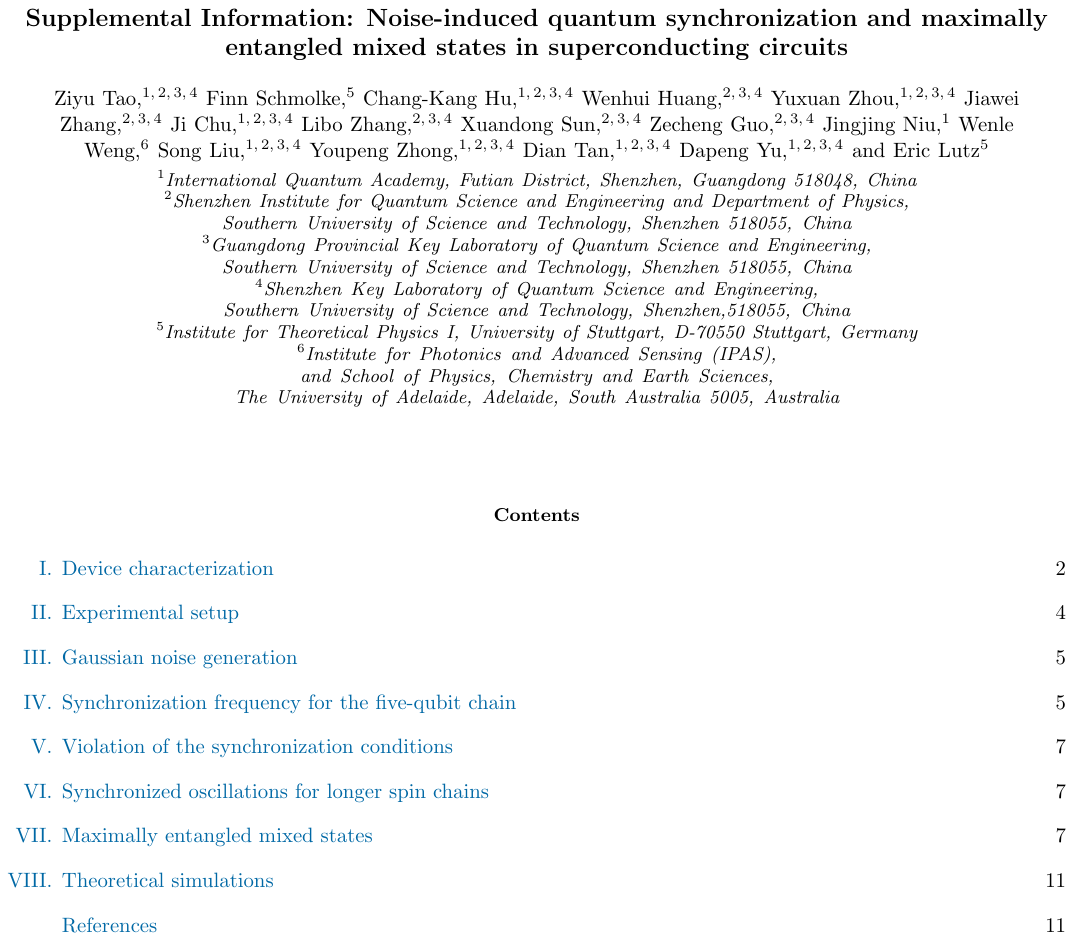}

\clearpage

\section{Device characterization}

\begin{figure}[!htbp]
	\centering
	\includegraphics[width=0.6\textwidth]{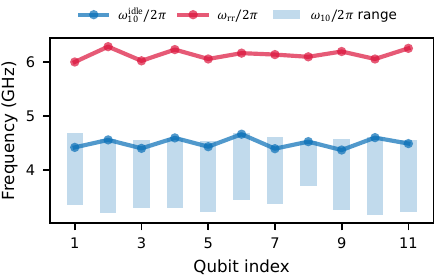}
	\caption{\label{qubit_para}
		Individual qubit frequencies in our setup for a linear chain of up to $N = 11$ qubits. Blue dots denote the idle (unmodified) frequency $\omega_{10}/2\pi$ and the blue bars denote the bandwidth of each qubit. Red lines with dots denote the readout resonator frequencies $\omega_{rr}/2\pi$.
	}
\end{figure}

\begin{figure}[!htbp]
	\centering
	\includegraphics[width=0.6\textwidth]{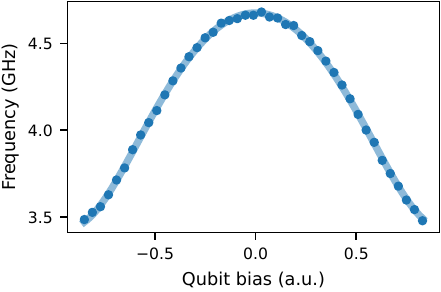}
	\caption{\label{qfreq_bias}
		\figtitle{
			Typical relation between the qubit frequency and the flux amplitude (qubit bias) applied through $Z$ control line.
		}
	}
\end{figure}

In this experiment, we use two asymmetric Josephson junctions with $E_{J1}/E_{J2}=3.0$ on the qubits, 
where $E_{J1}$ and $E_{J2}$ are the Josephson energies of the two junctions. 
The frequency (level spacing) $\omega_j$ of each qubit can be individually adjusted by varying the corresponding external flux through the $Z$ control line and ranges from approximately 3.2 GHz to 4.6 GHz.
In \rfig{qubit_para}, we show the qubit frequencies of the linear chain of up to $N = 11$.
In the main text, we focus on a consecutive set of five qubits as detailed below.
In \rfig{qfreq_bias} we plot a typical relation between the tunable qubit frequency and the amplitude of applied flux.

\begin{figure}[!htbp]
	\centering
	\includegraphics[width=0.7\textwidth]{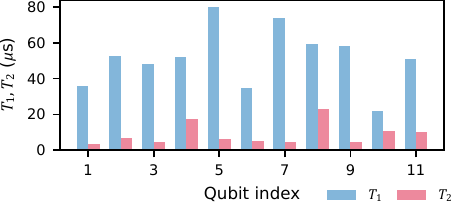}
	\caption{\label{qubit_T1T2} 
		{Qubit energy relaxation time $T_1$ and dephasing time $T_2$.}
	}
\end{figure}

Each qubit can be individually addressed and driven into the excited state by applying a microwave pulse through its $XY$ control line.
Figure \ref{qubit_T1T2} gives the energy relaxation ($T_1$) and dephasing time ($T_2$) of each qubit at its idling frequency. 
In the experiments involving 5, 8 and 11 working qubits, we choose these qubits from the index sets $\{4,5,6,7,8\}$, $\{4,5,6,\ldots ,11\}$ and $\{1,2,3,\ldots ,11\}$ respectively, where the frequency of the other unused qubits are largely detuned from the working qubits in the chosen index set of experiment, so the unused qubits do not influence the desired evolution of the working qubits ($5$, $8$ and $11$ qubits in each experiment).

\begin{figure}[!htbp]
	\centering
	\includegraphics[width=0.7\textwidth]{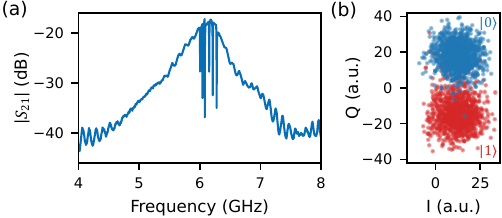}
	\caption{
		\figtitle{Qubit dispersive readout with Purcell filter. }
		\subfiglabel{a}
		Measured transmission spectrum of the Purcell filter.
		\subfiglabel{b}
		Single-shot dispersive readout for states $\vert 0\rangle, \vert 1\rangle$ in the quadrature (IQ) space. 
	}\label{s21_IQraw}
\end{figure}

\begin{figure}[!htbp]
	\centering
	\includegraphics[width=0.7\textwidth]{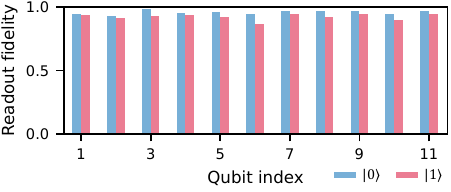}
	\caption{\label{qubit_readfid} 
		{Qubit readout fidelities for states $\vert 0\rangle$ and $\vert 1\rangle$.}
	}
\end{figure}

The dedicated readout resonator with frequency around 6.1 GHz is coupled to each qubit, 
where the state of the qubit can be deduced by measuring the state-dependent transmission of the readout resonator using the dispersive readout scheme.
To mitigate qubit relaxation we use Purcell filters, where each group of 6 readout resonators is coupled to a Purcell filter with the center frequency of 6.1 GHz, which can be used as a bandpass filter to impede microwave propagation at the qubit frequency and suppress the Purcell decay rate, 
as seen in its transmission spectrum in \rsubfig{s21_IQraw}{a}.
High-quality single-shot qubit dispersive readout for states $\ket{0}, \ket{1}$ can be achieved by using this Purcell filter, as shown in \rsubfig{s21_IQraw}{b}.
Figure \ref{qubit_readfid} displays the qubit readout fidelities, with an average state fidelity of $0.96$ for the $\ket{0}$ state, $0.92$ for the $\ket{1}$ state, respectively.

\begin{figure}[!tbhp]
	\centering
	\includegraphics[width=0.7\textwidth]{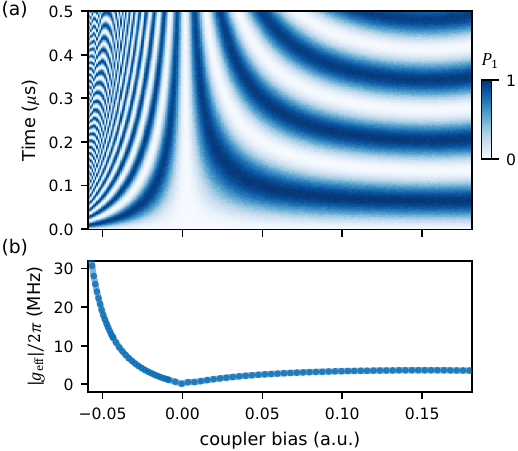}
	\caption{\label{iswap_cpa2d}
		\figtitle{Characterization of tunable coupling.}
		\subfiglabel{a}
		Vacuum Rabi oscillation between $Q_\mathrm{A}$ and $Q_\mathrm{B}$ at different coupler bias,
		where the colors denote the population $P_{1}$ of $Q_\mathrm{B}$.
		\subfiglabel{b}
		The tunable coupling strength $g_{\mathrm{eff}}$ extracted from the data of (a).
	}
\end{figure}

The nearest-neighbour coupling strength between qubits can be controlled by applying external flux on the corresponding coupler~\cite{Xu2020a_sm}.
As shown in \rfig{iswap_cpa2d}, 
we perform vacuum Rabi oscillation between the first excited states of $Q_\mathrm{A}$ and $Q_\mathrm{B}$ at different coupler bias to characterize the effective coupling strength between qubits,
where the coupling strength $g_{\mathrm{eff}}$ can be continuously adjusted from $+3.5$~MHz to approximately $-30$~MHz \cite{Yan2018}.

\section{Experimental setup}

\begin{figure}[!hbtp]
	\centering
	\includegraphics[width=0.65\textwidth]{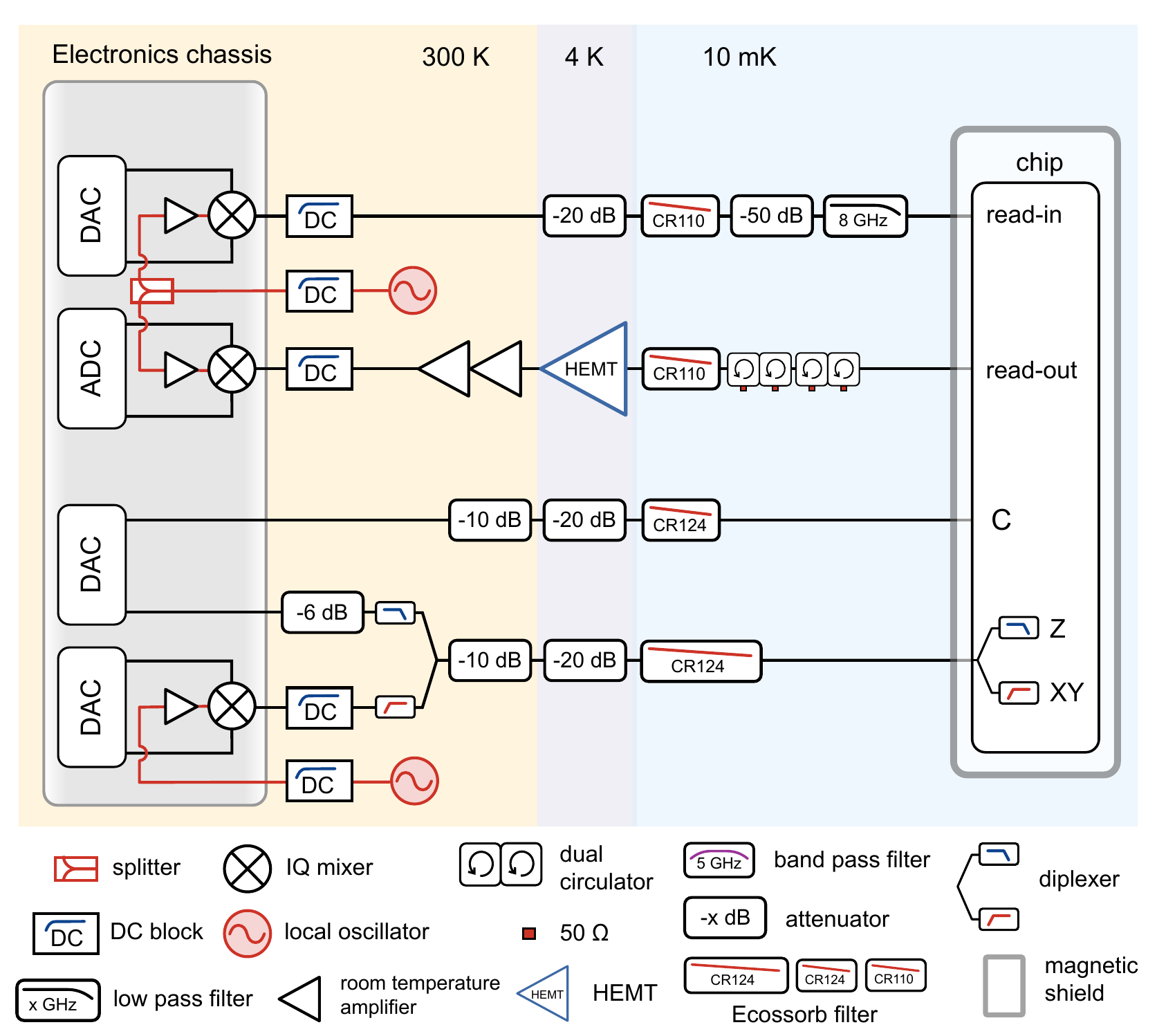}
	\caption{{ Room temperature and cryogenic wiring for the qubit control and readout.}
	}\label{wiring}
\end{figure}

Figure \ref{wiring} shows the room temperature and cryogenic wiring layout used in our experiments, which is similar to our previous work \cite{Tao2023}. 
In the room temperature part of our setup, we use custom made digital-to-analog converter (DAC) and analog-to-digital converter (ADC) circuit boards for qubit control and readout, respectively.
The control boards have dual-channel 14-bit vertical resolution DAC integrated circuits operating at 1~Gs/s driven by a field-programmable gate array (FPGA) chip, 
where each DAC analog output is filtered by a custom Gaussian low-pass filter with 250 MHz bandwidth to filter the clock feedthrough.
The DAC boards can generate nanosecond-length pulses for fast qubit $Z$ or coupler $C$ control, 
or modulate the in-phase and quadrature components of an IQ mixer for frequency up-conversion by using two channels, 
which provides several GHz frequency signals for qubit XY control and dispersive readout.
In the cryogenic wiring, 
two cryogenic circulators are inserted between the qubits chip and the cryogenic HEMT to isolate reflections as well as thermal noise emitted from the input of the cryogenic HEMT to the chip,
each control line is heavily attenuated and filtered at each temperature stage in the dilution refrigerator to minimize the impact on the qubit coherence while retaining controllability.

\section{Gaussian noise generation}

\begin{figure}[!htbp]
	\includegraphics[width=0.7\textwidth]{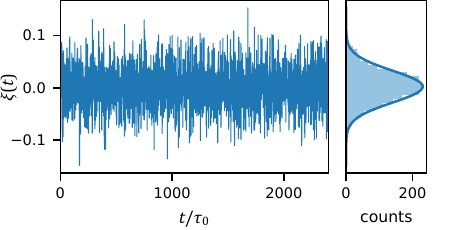}
	\caption{
		{Illustration of the pulse sequence injecting the white noise with time series $\xi(t)$.}
		The amplitude of white noise obeys the Gaussian distribution.
	}\label{noise_pulse_illu}
\end{figure}

To inject the artificial noise into a superconducting qubit, we use the method discussed in Ref.~\cite{Averin2016_sm}. 
We apply a series of pulse sequences in the experiments and treat the outputs as an ensemble of trajectories, whose average gives an effective open quantum system evolution for the qubits.
Each pulse sequence consists of successive 4-ns-wide square pulses, which are specifically designed to yield the desired noise spectrum.
The noise spectral density $S(\omega)$ is first discretized into the series $S(\omega_m)$, where $\omega_m = 2\pi m/(Nk\tau_0)$ ($m=0,1,2,\cdots ,Nk/2$), $\tau_0=4~\mathrm{ns}$, $N$ is the number of pulse sequences, $k$ is the maximum number of successive 4-ns-wide pulses in each sequence, and $k\tau_0$ is the time duration of each sequence. 
For Gaussian white noise, $S(\omega)$ is simply a constant.
We then multiply $\sqrt{S (\omega_m)}$ by a random phase factor $\theta_m$ which is randomly chosen from $0$ to $2\pi$, and perform the inverse Fourier transform of the frequency series 
$\{\sqrt{S(\omega_m)}\exp(i\theta_m)$  $,m=0,1,\cdots Nk/2\}$ 
to generate a discrete white noise time series $\xi(t)$ consisting of $Nk$ numbers.
The time series $\xi(t)$ is sliced into $N$ sections providing the individual noise trajectories, 
which are transformed to the amplitudes of square pulses as experimentally generated by the digital-to-analog converter (arbitrary waveform generator).
Figure \ref{noise_pulse_illu} illustrates a pulse sequence of white Gaussian noise.

\section{Synchronization frequency for the five-qubit chain}
\begin{figure}[!htbp]
	\centering
	\includegraphics[width=0.6\textwidth]{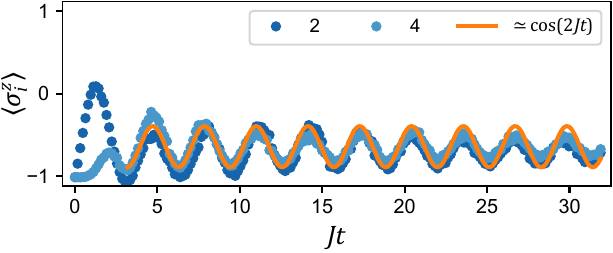}
	\caption{{
			The synchronized magnetization $\langle \sigma_j^z\rangle$ oscillates with the period of $2J$, which fits a cosine-like function $c_1\cos(2Jt)+c_2$ with constants $c_1,c_2$.
		}
	}\label{evo5q_cosine2Jt}
\end{figure}

The synchronization frequency $2J$ is determined by the coupling constant of the spin chain and not by the eigenfrequencies of the individual qubits. Figure \ref{evo5q_cosine2Jt} shows a fit the magnetizations of qubits $2-4$ for the five-qubit chain of the main text with the function $c_1\cos(2Jt)+c_2$.

\section{Violation of the synchronization conditions}

\begin{figure}[!htbp]
	\centering
	\includegraphics[width=0.55\textwidth]{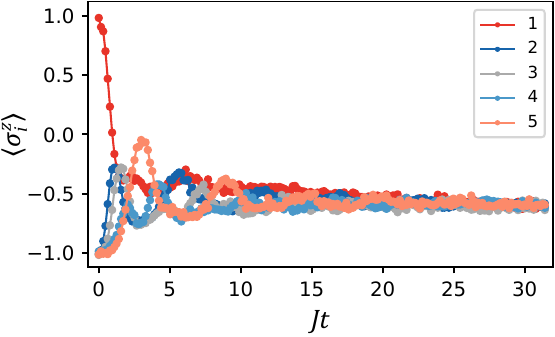}
	\caption{{Evolution of the five-qubit quantum XY chain with Gaussian white noise applied to the first site $V=\sigma^z_1$. The synchronization condition is not satisfied and the qubits reach a time-independent stationary state without oscillations (same parameters as in the main text).}
	}\label{supply_evo5q_site1}
\end{figure}

\begin{figure}[!htbp]
	\centering
	\includegraphics[width=0.55\textwidth]{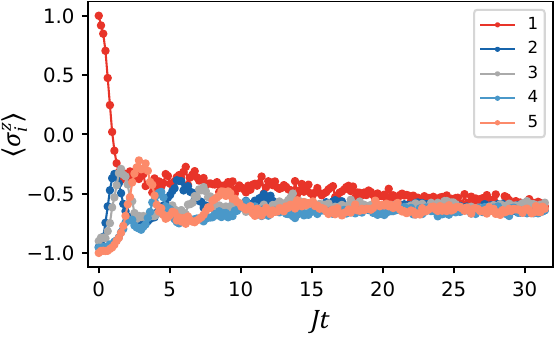}
	\caption{{Evolution of the five-qubit quantum XY chain with Gaussian white noise applied to the second site $V=\sigma^z_2$. The synchronization condition is not satisfied and the qubits reach a time-independent stationary state without oscillations (same parameters as in the main text).
		}
	}\label{supply_evo5q_site2}
\end{figure}

Synchronization with a single frequency is only possible in the $XY$ chain with noise if the synchronization conditions is satisfied (see main text).
When the conditions for transient synchronization are not obeyed, the quantum system will exhibit damped oscillations resulting in a time-independent steady state (no oscillations), cf. Ref.~\cite{Schmolke2022}.
In this section, we experimentally demonstrate this situation with the five-qubit example of the main text.
Figure \ref{supply_evo5q_site1} (\rfig{supply_evo5q_site2}) show the five-qubit evolution where  Gaussian white noise is applied to the first site $V=\sigma^z_1$ (second site $V=\sigma^z_2$).
As predicted by theory, we find no synchronization in this case.

\section{Synchronized oscillations for longer spin chains}

\begin{figure}[!htbp]
	\centering
	\includegraphics[width=0.6\textwidth]{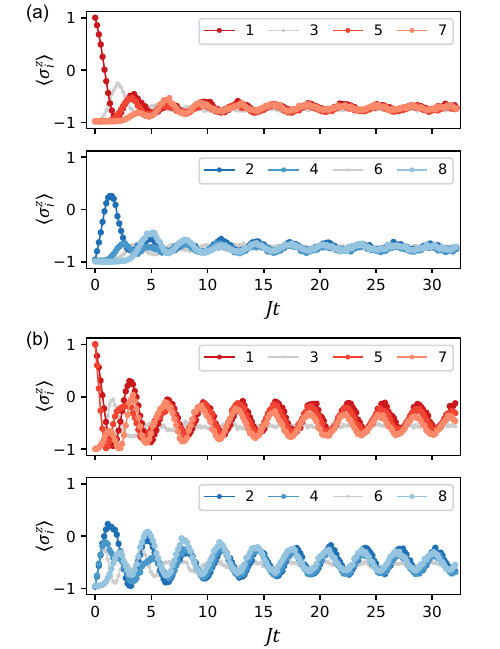}
	\caption{\label{fig_8qEvo}
		\subfiglabel{a}-\subfiglabel{b}
		Experimental demonstration of noise-induced synchronization in a $XY$ model with $N=8$ qubits.
		Measured magnetization $\langle \sigma_j^z\rangle$ in the $8$-qubit chain.
		The system is initially prepared in the state 
		\subfiglabel{a} $\vert \Psi(t=0)\rangle $ $= \vert 1\rangle_1 $, 
		\subfiglabel{b} $\vert \Psi(t=0)\rangle = \vert 1\rangle_1 \otimes \vert 1\rangle_5$.
	}
\end{figure}

\begin{figure}[!htbp]
	\centering
	\begin{tikzpicture}
		\node (a) [label={[label distance=-0.6 cm]150: \textbf{\textsf{(a)}}}] at (0,0) {\includegraphics{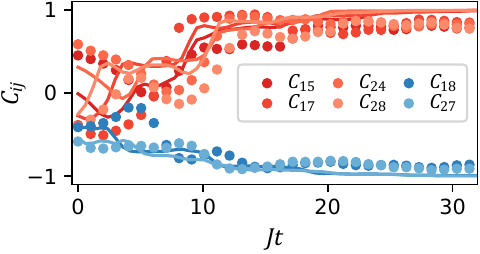}};	
		\node (b) [label={[label distance=-0.6 cm]150: \textbf{\textsf{(b)}}}] at (8.5,0) {\includegraphics{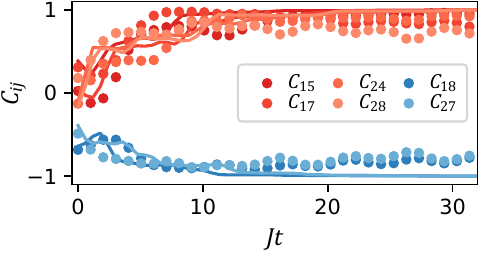}}; 
	\end{tikzpicture}
	\caption{{
			Pearson correlation coefficients extracted from the evolution of $\langle \sigma_j^z\rangle$ in a 8-qubit chain (dots) and numerical simulation (solid line) with the initial state
			\subfiglabel{a} $\vert \Psi(t=0)\rangle = $ $ \vert 1\rangle_1 $, 
			\subfiglabel{b} $\vert \Psi(t=0)\rangle = $ $ \vert 1\rangle_1 \otimes \vert 1\rangle_5$.
		}
	}\label{fig_evo8q_pearson}
\end{figure}

\begin{figure}[!htbp]
	\centering
	\includegraphics[width=0.5\textwidth]{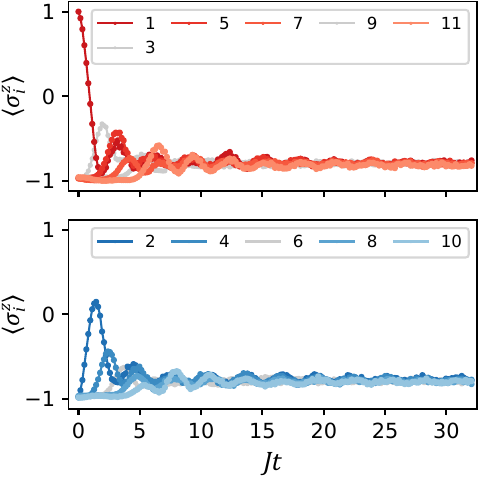}
	\caption{\label{fig_11qEvo_1ex}
		Experimental demonstration of noise-induced synchronization in a $XY$ model with $N=11$ qubits.
		Measured magnetization $\langle \sigma_j^z\rangle$ in a $11$-qubit chain
		with the initial state $\vert \Psi(t=0)\rangle$ $= \vert 1\rangle_1 $.
	}
\end{figure}

\begin{figure}[!htbp]
	\centering
	\includegraphics[width=0.5\textwidth]{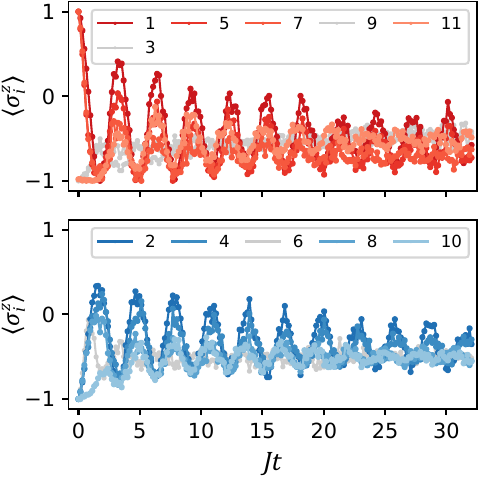}
	\caption{\label{fig_11qEvo_3ex}
		Experimental demonstration of noise-induced synchronization in a $XY$ model with $N=11$ qubits.
		Measured magnetization $\langle \sigma_j^z\rangle$ in a $11$-qubit chain
		with the initial state $\vert \Psi(t=0)\rangle$ $= \ket{1}_1\otimes \ket{1}_5 \otimes \ket{1}_7$.
	}
\end{figure}

In this section, we demonstrate the synchronized magnetization $\langle \sigma_j^z\rangle$ in a chain of length $N=8$ and $N=11$. 
For $N=8$, Gaussian white noises are locally applied on the third and sixth site $u = 3,6$ with the reduced noise amplitude $\gamma \approx 0.5$.
Figure \ref{fig_8qEvo} gives the measured magnetization $\langle \sigma_j^z\rangle$ in a 8-qubit chain, for the initial state with a single excitation $\vert \Psi(t=0)\rangle = \vert 1\rangle_1 $ and two excitations $\vert \Psi(t=0)\rangle = \vert 1\rangle_1 \otimes \vert 1\rangle_5$, as shown in \rsubfig{fig_8qEvo}{a} and \rsubfig{fig_8qEvo}{b}.
The evolution of $\langle \sigma_j^z\rangle$ shows a synchronous behavior for $\langle \sigma_{1,5,7}^z\rangle$ and $\langle \sigma_{2,4,8}^z\rangle$.
Figure \ref{fig_evo8q_pearson} shows the Pearson correlation coefficients extracted from \rfig{fig_8qEvo}.
For $N=11$, Gaussian white noises are locally applied on the sites $u = 3,6,9$ with the reduced noise amplitude $\gamma \approx 0.3$.
Figure \ref{fig_11qEvo_1ex} and \rfig{fig_11qEvo_3ex} give the measured magnetization $\langle \sigma_j^z\rangle$ in a 11-qubit chain, for the initial state with single excitation $\vert \Psi(t=0)\rangle = \vert 1\rangle_1 $ and three excitations $\vert \Psi(t=0)\rangle = \vert 1\rangle_1 \otimes \vert 1\rangle_5 \otimes \vert 1\rangle_7$, respectively.
In all cases, all eigenmodes are quickly suppressed until only a single one survives giving rise to synchronized oscillations across the entire system at a single frequency.

\section{Maximally entangled mixed states}
The quantum $XY$ chain subject to noise on the third site possess a decoherence-free subspace that supports a single frequency, the synchronization frequency $\hbar\Lambda_{kl} = 2J$ \cite{Schmolke2022}.
In this section, we theoretically show that, as long as the synchronization conditions are satisfied (see main text), the two states that give rise to the decoherence-free eigenmode, lead to maximally entangled mixed states between the end-spins of the chain for arbitrary system size $N$.

To simplify the calculations, we can either perform a Jordan-Wigner transformation \cite{Schmolke2022} or only consider the subspace of a single excitation (since, even in the presence of noise, the total magnetization is conserved).
In both cases, we obtain a von Neumann equation with an effective tridiagonal Hamiltonian \cite{Schmolke2022}
\begin{align}
	\Omega/\hbar = \mathrm{diag}(J;2\omega;J).
\end{align}
The eigenstates of this matrix are given by \cite{Noschese2013} 
\begin{align}
	\ket{\varphi_k} = 
	\left(\sin(\frac{\pi k}{N+1}),\sin(\frac{2\pi k}{N+1}),\ldots,\sin(\frac{N\pi k}{N+1})\right)^\mathrm{T} \sqrt{\frac{2}{N+1}},
\end{align}
The two decoherence-free states that give rise to non-decaying oscillations in the long time limit have already been identified in Ref.~\cite{Schmolke2022} and are given by $k = (N+1)/3$ and $l = 2(N+1)/3$, yielding
\begin{align}
	\ket{\varphi^\mathrm{s}_k} &=
	\left(\sin(\frac{\pi}{3}),\sin(\frac{2\pi}{3}),\ldots,\sin(\frac{N\pi}{3})\right)^\mathrm{T} \sqrt{\frac{2}{N+1}}, \\
	\ket{\varphi^\mathrm{s}_l} &=
	\left(\sin(\frac{2\pi}{3}),\sin(\frac{4\pi}{3}),\ldots,\sin(\frac{2N\pi}{3})\right)^\mathrm{T} \sqrt{\frac{2}{N+1}}.
\end{align}
Transforming back into original Hilbert space, these states become
\begin{align}
	\ket{v^\mathrm{DFS}_1} &= 
	\sqrt{\frac{2}{N+1}} \sum_{n=1}^N \sin(\frac{n\pi}{3}) \ket{1}_n, \\
	\ket{v^\mathrm{DFS}_2} &= 
	\sqrt{\frac{2}{N+1}} \sum_{n=1}^N \sin(\frac{2n\pi}{3}) \ket{1}_n.
\end{align}
Now, performing the partial trace over all subsystems except the edge qubits, we arrive at
\begin{align}
	\tr_{[2,N-1]}(\dyad{v^\mathrm{DFS}_1})
	=& 
	\frac{2}{N+1} \sum_{m,n,k} \sin(\frac{n\pi}{3})\sin(\frac{m\pi}{3}) \tr_{[2,N-1]}(\ket{1}_n \bra{1}_m),\\
	=& \frac{2}{N+1} \bigg[\frac{2(N-2)}{3}\sin(\frac{\pi}{3})^2 \dyad{00} + \sin(\frac{N\pi}{3})^2 \dyad{01}\\
	&+ \sin(\frac{(N-1)\pi}{3})^2 \dyad{10}\\
	&+ \sin(\frac{N\pi}{3})\sin(\frac{(N-1)\pi}{3})(\dyad{01}{10}+\dyad{10}{01})\bigg].
\end{align}
Since, $\sin(n\pi/3)^2 = 3/4$ for $n \in \mathbb{N}$, this two-qubit mixed state can be conveniently expressed in matrix form 
\begin{align}
	\label{s11}
	\rho_{1,N} = \frac{3}{2N+2}
	\begin{pmatrix}
		0 & 0 & 0 & 0 \\
		0 & 1 & -1 & 0 \\
		0 & -1 & 1 & 0 \\
		0 & 0 & 0 & \frac{2N-4}{3}
	\end{pmatrix} =M.
\end{align}
States of this form belong to the class of maximally entangled mixed states and can be parametrized as \cite{Ishizaka2000}
\begin{align}
	\rho = p_1 \dyad{\Psi^-} + p_2 \dyad{00} + p_3 \dyad{\Psi^+} + p_4 \dyad{11},
\end{align}
with $\sum_k p_k = 1$ where $\ket{\Psi^-} = (\ket{01} - \ket{10})/\sqrt{2}$ and $\ket{\Psi^+} = (\ket{01} + \ket{10})/\sqrt{2}$ are the usual Bell states.
For the above states, we have 
\begin{align}
	p_1 = \frac{3}{N+1}, \quad p_2 = \frac{N-2}{N+1}, \quad p_3 = p_4 = 0.
\end{align}
The purity, $\tr(\rho_{1,N}^2) = (9+(N-2)^2)/(N+1)^2$, and the concurrence, $C(\rho_{1,N}^2) = 3/(N+1)$, immediately follow \cite{Ishizaka2000}. The concurrence is nonzero for all finite $N$.

Since, $\ket{v^\mathrm{DFS}_{1,2}}$ are decoherence-free subspaces, they are in the kernel of the Liouvillian, $\mathcal{L} \dyad{v^\mathrm{DFS}_{1,2}} = 0$, and thus belong to the invariant states if they are excited by the initial condition.
In the main text, the initial condition for noise-induced synchronization is $\ket{\Psi(0)} = \ket{1}_1$.
Both states have a non-zero overlap $|\braket{v^\mathrm{DFS}_{1,2}}{\Psi(0)}|^2 = 3/(2N+2)$ with the initial condition.
Consequently, the stationary state contains the maximally entangled mixed state contribution $3/(2N+2)(\dyad{v^\mathrm{DFS}_1} + \dyad{v^\mathrm{DFS}_2})$.

\begin{figure}[!htbp]
	\centering
	\includegraphics[width=0.6\textwidth]{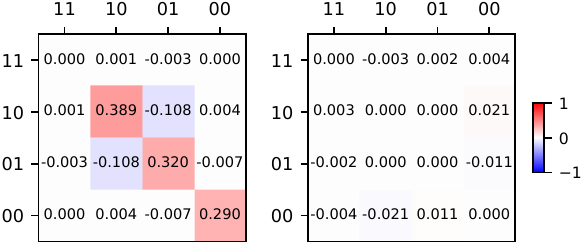}
	\caption{{
			The real (left) and imaginary (right) parts of the experimentally extracted density matrix 
			$\rho_{15}$ of the two end qubits in the synchronized regime of $5$-qubit experiment at $Jt =3\pi$.
		}
	}\label{rho15_Mstate_exp}
\end{figure}

Real and imaginary parts of the  experimentally reconstructed state $\rho_{15}$ at $Jt = 3\pi$ (corresponding to  Fig.~1f of the main text) are presented in Fig.~\ref{rho15_Mstate_exp}. The real part has the form of a maximally entangled mixed state (the fidelity with Eq.~\eqref{s11} is 99.3$\%$), while the imaginary part is basically zero.

\section{Theoretical simulations}

The lattice model corresponding to our experiments can be described by the Hamiltonian of a one-dimensional quantum $XY$ chain of $N$ spins \cite{Schmolke2022}
\begin{equation}
	H_0 = \frac{J\hbar}{2} \sum_{j=1}^{N-1} 
	\left(\sigma_j^x \sigma_{j+1}^x + \sigma_j^y \sigma_{j+1}^y\right)
	+ \sum_{j=1}^N \hbar \omega_j \sigma_j^z
\end{equation}
where $\sigma_j^{x,y,z}$ are the local Pauli operators acting on site $j$, 
$J$ is nearest-neighbor coupling strength, and $\omega$ is the level spacing of the qubits.
It has been shown that locally adding noise to the level spacing of a single qubit is sufficient to induce synchronization in a chain of arbitrary length $N$, provided a synchronization condition is satisfied.
Concretely, we introduce a white noise process $\xi(t)$ with zero mean $\langle \xi(t)\rangle$ and auto-correlation $\langle \xi(t) \xi(t^\prime) \rangle = \Gamma \delta(t-t^\prime)$ via a Hermitian operator $V = \sigma^z_u$ (acting locally on site $u$), resulting in an exact Lindblad master equation for the evolution of the density matrix \cite{Schmolke2022}
\begin{align}
	\dot{\rho} = -\frac{i}{\hbar} [H_0,\rho] + \Gamma \left(V\rho V^\dagger - \frac{1}{2}\{V^\dagger V,\rho\}\right),
	\label{eq:me}
\end{align}
where $\Gamma$ denotes the strength of the noise.
In order to satisfy the synchronization condition (see main text), white noise is applied to the third site $u = 3$ with $V = \sigma^z_3$.
We numerically solve the master equation \eqref{eq:me} for $\rho$ as a function of time from which we then compute the Pearson correlator $C_{ij}$, concurrence $\mathcal{C}(\rho_{15})$ and the maximally entangled mixed state fidelity $F(M,\rho_{15})$, corresponding to the theoretical curves in the main text.

\end{document}